# Can a balance of electric and gravitational forces be achieved? Remark to "Retraction: Conservative relativity principle and energy-momentum conservation in a superimposed gravitational and electric field" by Alexander Kholmetskii and Tolga Yarman

Alexander Kholmetskii[1] and Tolga Yarman[2]

[1]Department of Physics, Belarus State University, Minsk, Belarus, e-mail: alkholmetskii@gmail.com
[2]Okan University, Istanbul, Turkey & Savronik, Eskisehir, Turkey

We explain our strong disagreement with the statement about "several scientific errors" in our paper and highlight the validity of our approach, which had been already confirmed in the well-known experiments by Millikan.

Recently, Canadian Journal of Physics (CJP) published the retraction with respect to our paper [1], stating that "this paper contains several scientific errors" [2]. We absolutely disagree with this statement, and continue to be sure that the paper is fully correct. In any case, the readers have the right to be aware about the claimed "several scientific errors" (which are not indicated in [2]) and, on the other hand, we have further the responsibility to defend our scientific reputation.

The paper has been submitted in April, 2017, and accepted for publication on 15 June, 2017. However, on 17 July, 2017 we were informed about the reverse decision with respect to this paper, with the following comment of one of the associate editors: "The authors use the example of electromagnetic and gravitational forces cancelling. This is impossible. They have different spins and different polarizations. Their physics is wrong."

In spite of obvious non-accuracy of this comment (a force itself cannot be characterized by polarization and spin), what is claimed by the associate editor, is understandable: he comes to mean, due to the substantial difference between electromagnetic and gravitation interactions, a dynamic balance of electromagnetic and gravitational forces, in general, cannot be achieved, and we do not argue against this statement. However, for the *static* electric and gravitation fields (which is just the case for the paper [1]), their balance can be actually realized, and this statement is fully supported by the well-known experimental facts (e.g., the experiments by Millikan [3], where the additional presence of the Stokes force did not play a principal role).

Indeed, we remind that within the framework of the general theory of relativity, the gravitational force is defined as the covariant derivative of three-momentum of moving object with respect to its proper time, and in the static (or stationary) case it has the form [4]

$$\boldsymbol{F}_g = \gamma mc^2 \left( -\nabla \ln \sqrt{g_{00}} + \frac{\sqrt{g_{00}}}{c} \left( \boldsymbol{v} \times (\nabla \times \boldsymbol{g}) \right) \right), \qquad (1)$$

where $\gamma$ is the Lorentz factor of the object with the rest mass $m$ measured by a local observer, $\boldsymbol{v}$ is the velocity of this object for a local observer, and the components of vector $\boldsymbol{g}$ are equal to $g_i = -g_{0i}/g_{00}$ ($i=1…3$), $g_{\alpha\beta}$ being the metric tensor ($\alpha, \beta=0…3$). In a static gravitation field the metric coefficients $g_{0i}$ are equal to zero, and eq. (1) takes the simpler form

$$\boldsymbol{F}_g = -\gamma mc^2 \nabla \ln \sqrt{g_{00}}. \qquad (2)$$

Further, in the one-dimensional case considered for simplicity in ref. [1], where the coefficient $g_{00}$ depends only on the $x$-coordinate, we derive that the gravitational force has the single non-vanishing component collinear to the axis $x$:

$$(F_g)_x = -\gamma mc^2 \frac{\partial}{\partial x} \ln \sqrt{g_{00}} = -\frac{\gamma mc^2}{2g_{00}} \frac{dg_{00}}{dx}. \qquad (3)$$

Next, we consider the electric force on the charged particle, which is defined via the motional equation of charged particle in an electromagnetic field in the presence of gravity (see, e.g. [4, 5]):

$$mc\frac{Du^\mu}{ds} = \frac{e}{c}F^{\mu\nu}u_\nu, \qquad (4)$$

where $Du^\mu/ds$ stands for covariant derivative, $ds$ is the space-time interval, $F^{\mu\nu}$ is the tensor of electromagnetic field, and $u^\mu$ is the four-velocity of particle.

In the case, where the electric field has a single non-vanishing $x$-component $E$, and the magnetic field is absent, we get a single non-vanished component of the electromagnetic energy-momentum tensor [4]

$$F^{10} = E/\sqrt{g_{00}}.$$

Here the value of electric field $E$ and the corresponding scalar potential $\varphi$ (so that $E = -\partial\varphi/\partial x$) are defined in the absence of gravity. Involving the known expression for the zeroth component of four-velocity in gravitation field, $u_0 = \gamma/\sqrt{g_{00}}$ [6], we derive the electric force, acting on the particle, which has a single non-vanishing $x$-component, being collinear to the axis $x$:

$$(F_e)_x = -\frac{e}{g_{00}}\frac{\partial\varphi}{\partial x}. \qquad (5)$$

Thus, eqs. (3) and (5) indicate that in a static superimposed gravitational and electric field, we can always realize the configuration, where the gravitational and electric forces are collinear to each other. By this way, we demonstrate the necessary condition for the achievement of a balance between these forces.

The sufficient condition for the realization of this balance is expressed by the equality

$$(F_g)_x + (F_e)_x = 0,$$

which, via eqs. (3) and (5), yields

$$-\frac{\gamma mc^2}{2e}\frac{dg_{00}}{dx} = \frac{\partial\varphi}{\partial x}. \qquad (6)$$

This is the basic equation used in our analysis of ref. [1], and we are strongly sure that the derivation of this equation is fully correct.

Thus, we firmly refuse the statement of [2] about "several scientific errors" in our paper [1]. Moreover, eq. (6) completely agrees with the known experimental fact already mentioned above – the experiment by Millikan [3] – where the balance between the static electric and gravitational forces played the principal role in the performance of this historically vital experiment, aimed to determine the value of elementary charge.

In our paper [1], we analyze some important implications of eq. (6), which in particular show that the balance between the electric and gravitational forces occurs to be velocity-dependent due to the presence of $\gamma$-factor in its left side, corresponding to the gravitational force component. Of course, in the experiments by Millikan, where the velocities of charged oil drops were very small, this dependence could be totally ignored. However, with the increase of $v$, when we achieve a low velocity relativistic limit (corresponding to the accuracy of calculations $(v/c)^2$), the dependence of the balance between electric and gravitational forces on the velocity of particle creates serious difficulties in the implementation of the total energy-momentum conservation law [1], which are eliminated only with the adoption of conservative relativity principle (CRP), which we recently advanced [7].

Thus, we believe that our paper [1] is fully correct from the physical viewpoint, and it contains some results of the principal importance. In these conditions, we actually regret with respect to the decision of CJP to retract this paper, which we believe is interesting and important.